# Quantization of the free electromagnetic field implies quantization of its angular momentum

### A. M. Stewart

*Emeritus Faculty, The Australian National University,
Canberra, ACT 0200, Australia.*
http://grapevine.net.au/~a-stewart/index.html

### Abstract

It is shown that when the gauge-invariant Bohr-Rosenfeld commutators of the free electromagnetic field are applied to the expressions for the linear and angular momentum of the electromagnetic field interpreted as operators then, in the absence of electric and magnetic charge densities, these operators satisfy the canonical commutation relations for momentum and angular momentum. This confirms their validity as operators that can be used in quantum mechanical calculations of angular momentum.

### 1. Introduction

The free electromagnetic field was first quantized in a plane wave basis in the Coulomb gauge and later in the Lorenz gauge [1,2,3,4]. The four components of the electromagnetic potential were promoted to operators and the commutation relations between them determined by analogy with those of the simple harmonic oscillator. The commutations relations are different in each gauge [3]. The quanta of the electromagnetic field are known as photons and the states that the operators act on are Fock states of photons (not necessarily plane waves) [5]. By taking appropriate combinations of the derivatives of the potentials, Bohr and Rosenfeld [6] obtained the gauge-invariant equal time commutation relations between the field operators $\mathbf{E}(\mathbf{x},t)$ and $\mathbf{B}(\mathbf{y},t)$

$$[E^i(\mathbf{x},t), B^j(\mathbf{y},t)] = -\frac{i\hbar}{\varepsilon_0}\varepsilon^{ijk}\frac{\partial}{\partial x^k}\delta(\mathbf{x}-\mathbf{y}) \quad ; \quad (1)$$

all other commutators are zero. SI units are used, $\varepsilon$ is the Levi-Civita symbol and the square brackets $[A,B]$ represent the commutator $AB - BA$ where $A$ and $B$ are two operators. Bohr and Rosenfeld used equation (1) to investigate the effect of the quantum uncertainty principle on the measurement of field strength. Because we will be dealing solely with operators that act at equal time $t$, we drop the time parameter. The linear momentum $\mathbf{P}$ and angular momentum $\mathbf{J}$ of the classical electromagnetic field, are given by such equal-time gauge-invariant expressions [7]

$$\mathbf{P} = \varepsilon_0 \int d^3x\, \mathbf{E}(\mathbf{x}) \times \mathbf{B}(\mathbf{x}) \quad (2)$$

and

$$\mathbf{J} = \varepsilon_0 \int d^3x\, \mathbf{x} \times [\mathbf{E}(\mathbf{x}) \times \mathbf{B}(\mathbf{x})] \quad . \quad (3)$$

The canonical quantization of the motion of a free *particle* leads to commutation relations such as [2]

$$[P^i, P^j] = 0 \quad , \quad (4)$$

$$[P^i, J^j] = i\hbar \varepsilon^{ijk} P^k \quad , \quad (5)$$

$$[J^i, J^j] = i\hbar \varepsilon^{ijk} J^k \quad , \quad (6)$$

where the *P* and *J* are the components of the linear and angular momentum of the particle.

The question asked in this paper is whether the operators (2) and (3) for the momentum and angular momentum of the electromagnetic field lead, when the commutation relations (1) are used, to the commutation relations (4-6)? We will find that, in the absence of electric and magnetic charge densities, they do, thereby confirming that that (2) and (3) are valid quantum operators for these properties of the free electromagnetic field. This is done without invoking mode expansions of the fields [10]. In sections **2**, **3**, **4** of the paper we verify the commutation relations (4-6). In appendix A we confirm that the momentum density of the fields may be written as being proportional to either **ExB** or - **BxE** except in the presence of singularities of the field. In appendix B the most general expressions for the basis vectors in real space used in the quantization of the free field in the Coulomb gauge are given.

**2. Momentum-momentum commutators**
From equation (2) we have

$$[P^i, P^j]/\varepsilon_0^2 = \sum_{k,l,m,n} \varepsilon^{ikl} \varepsilon^{jmn} \int d^3x \int d^3y [E^k(\mathbf{x}) B^l(\mathbf{x}), E^m(\mathbf{y}) B^n(\mathbf{y})] \quad . \quad (7)$$

The expansion of the commutator in the integrand is

$$[E^k(\mathbf{x}) B^l(\mathbf{x}), E^m(\mathbf{y}) B^n(\mathbf{y})] = E^m(\mathbf{y})[E^k(\mathbf{x}), B^n(\mathbf{y})] B^l(\mathbf{x}) - E^k(\mathbf{x})[E^m(\mathbf{y}), B^l(\mathbf{x})] B^n(\mathbf{y}) \quad . \quad (8)$$

Putting (8) into (7) with the commutation relation (1), we arrive at

$$[P^i, P^j]/\varepsilon_0^2 = \frac{i\hbar}{\varepsilon_0} \sum_{k,l,m,n,s} \varepsilon^{ikl} \varepsilon^{jmn} \int d^3x \int d^3y [-\varepsilon^{kns} E^m(\mathbf{y}) B^l(\mathbf{x}) + \varepsilon^{mls} E^k(\mathbf{x}) B^n(\mathbf{y})] \frac{\partial}{\partial x^s} \delta(\mathbf{x} - \mathbf{y}) \quad . \quad (9)$$

In obtaining this result, we have used the algebraic identity

$$\frac{\partial}{\partial y^k} \delta(\mathbf{y} - \mathbf{x}) = -\frac{\partial}{\partial x^k} \delta(\mathbf{y} - \mathbf{x}) \quad . \quad (10)$$

Next, summing over the repeated indices of the product of Levi Civita symbols as

$$\sum_k \varepsilon^{ikl} \varepsilon^{kns} = \delta_{ln} \delta_{is} - \delta_{ls} \delta_{in} \quad , \quad (11)$$

we obtain

$$[P^i, P^j]/\varepsilon_0 = -i\hbar \sum_{k,l,m,n,s} \varepsilon^{jmn} \int d^3x \int d^3y$$
$$\times [(\delta_{ln}\delta_{is} - \delta_{ls}\delta_{in})E^m(\mathbf{y})B^l(\mathbf{x}) - (\delta_{is}\delta_{km} - \delta_{im}\delta_{ks})E^k(\mathbf{x})B^n(\mathbf{y})]\frac{\partial}{\partial x^s}\delta(\mathbf{x}-\mathbf{y})$$ (12)

and hence

$$[P^i, P^j]/\varepsilon_0 = -i\hbar \sum_{k,m,n} \int d^3x \int d^3y \{\varepsilon^{jmn}[E^m(\mathbf{y})B^n(\mathbf{x}) - E^m(\mathbf{x})B^n(\mathbf{y})]\frac{\partial}{\partial x^i}\delta(\mathbf{x}-\mathbf{y})$$
$$- [\varepsilon^{ijm}E^m(\mathbf{y})B^k(\mathbf{x}) + \varepsilon^{ijn}E^k(\mathbf{x})B^n(\mathbf{y})]\frac{\partial}{\partial x^k}\delta(\mathbf{x}-\mathbf{y})\}$$ (13)

The first term of (13) in square brackets vanishes, as can be seen by exchanging **x** and **y** and using (10). In second term in the second square brackets of (13) let $n \to m$ to give

$$[P^i, P^j]/\varepsilon_0 = i\hbar \sum_{k,m} \int d^3x \int d^3y \varepsilon^{ijm}[E^m(\mathbf{y})B^k(\mathbf{x}) + E^k(\mathbf{x})B^m(\mathbf{y})]\frac{\partial}{\partial x^k}\delta(\mathbf{x}-\mathbf{y}).$$ (14)

Now apply the algebraic identity

$$\int d^3x \int d^3y\, g(\mathbf{y})f(\mathbf{x})[\frac{\partial}{\partial x^i}\delta(\mathbf{x}-\mathbf{y})] = -\int d^3x\, g(\mathbf{x})[\frac{\partial}{\partial x^i}f(\mathbf{x})] = \int d^3x[\frac{\partial}{\partial x^i}g(\mathbf{x})]f(\mathbf{x})$$ (15)

to (14) to get

$$[P^i, P^j]/\varepsilon_0 = -i\hbar \sum_{k,m} \int d^3x\, \varepsilon^{ijm}[E^m(\mathbf{x})\frac{\partial B^k(\mathbf{x})}{\partial x^k} + \frac{\partial E^k(\mathbf{x})}{\partial x^k}B^m(\mathbf{x})].$$ (16)

The index $k$ is summed over $x, y, z$: if there are no electric or magnetic densities of charge, then both div**B** and div**E** are zero. Both terms of (16) therefore vanish and the relation (4) is proved.

**3. Momentum-angular momentum commutators**
The commutator to be determined is

$$[J^i, P^j] = \varepsilon_0^2 \sum_{k,l,n} \varepsilon^{jkl} \int d^3x \int d^3y\, x^n [E^i(\mathbf{x})B^n(\mathbf{x}) - E^n(\mathbf{x})B^i(\mathbf{x}), E^k(\mathbf{y})B^l(\mathbf{y})].$$ (17)

or

$$[J^i, P^j] = \varepsilon_0^2 \sum_{k,l,n} \varepsilon^{jkl} \int d^3x \int d^3y\, x^n \{[E^i(\mathbf{x})B^n(\mathbf{x}), E^k(\mathbf{y})B^l(\mathbf{y})] - [E^n(\mathbf{x})B^i(\mathbf{x}), E^k(\mathbf{y})B^l(\mathbf{y})]\}.$$ (18)

From (8) this becomes

$$[J^i, P^j] = \varepsilon_0^2 \sum_{k,l,n} \varepsilon^{jkl} \int d^3x \int d^3y \, x^n \{E^k(\mathbf{y})[E^i(\mathbf{x}), B^l(\mathbf{y})]B^n(\mathbf{x}) - E^i(\mathbf{x})[E^k(\mathbf{y}), B^n(\mathbf{x})]B^l(\mathbf{y})$$
$$- E^k(\mathbf{y})[E^n(\mathbf{x}), B^l(\mathbf{y})]B^i(\mathbf{x}) + E^n(\mathbf{x})[E^k(\mathbf{y}), B^i(\mathbf{x})]B^l(\mathbf{y})\} \quad , \quad (19)$$

and putting in (1) we get

$$[J^i, P^j] = i\hbar\varepsilon_0 \sum_{k,l,n,s} \varepsilon^{jkl} \int d^3y \int d^3x \, x^n \frac{\partial}{\partial x^s} \delta(\mathbf{x}-\mathbf{y})$$
$$\{-\varepsilon^{ils} E^k(\mathbf{y})B^n(\mathbf{x}) - \varepsilon^{kns} E^i(\mathbf{x})B^l(\mathbf{y}) + \varepsilon^{nls} E^k(\mathbf{y})B^i(\mathbf{x}) + \varepsilon^{kis} E^n(\mathbf{x})B^l(\mathbf{y})\} \quad . \quad (20)$$

Using relations such as (11) we find

$$[J^i, P^j] = -i\hbar\varepsilon_0 \sum_{n,s} \int d^3y \int d^3x \, x^n \frac{\partial}{\partial x^s} \delta(\mathbf{x}-\mathbf{y})$$
$$\{\delta_{ij}[E^n(\mathbf{x})B^s(\mathbf{y}) - E^s(\mathbf{y})B^n(\mathbf{x})] + \delta_{nj}[E^s(\mathbf{y})B^i(\mathbf{x}) - E^i(\mathbf{x})B^s(\mathbf{y})]$$
$$+ \delta_{js}[-E^n(\mathbf{x})B^i(\mathbf{y}) + E^i(\mathbf{y})B^n(\mathbf{x}) + E^i(\mathbf{x})B^n(\mathbf{y}) - E^n(\mathbf{y})B^i(\mathbf{x})]\} \quad . \quad (21)$$

Of the eight terms above, the first four vanish because of div**B** = 0 and div**E** = 0, using (15). The last four terms are listed below as integrands of $d^3x$, using (15) and putting $\mathbf{y} = \mathbf{x}$:

$$E^n(\mathbf{y})[x^n B^i(\mathbf{x})] \rightarrow -E^n(\mathbf{y})[\delta_{jn} + x^n \frac{\partial}{\partial x^j}]B^i(\mathbf{x})$$
$$E^n(\mathbf{x})x^n[B^i(\mathbf{y})] \rightarrow +E^n(\mathbf{x})x^n \frac{\partial}{\partial x^j} B^i(\mathbf{y})$$
$$-E^i(\mathbf{y})[x^n B^n(\mathbf{x})] \rightarrow +E^i(\mathbf{y})[\delta_{jn} + x^n \frac{\partial}{\partial x^j}]B^n(\mathbf{x}) \quad . \quad (22)$$
$$-E^i(\mathbf{x})x^n[B^n(\mathbf{y})] \rightarrow -E^i(\mathbf{x})x^n \frac{\partial}{\partial x^j} B^n(\mathbf{y})$$

The terms involving derivatives cancel. What is left is

$$[J^i, P^j] = i\hbar\varepsilon_0 \int d^3x [E^i(\mathbf{x})B^j(\mathbf{x}) - E^j(\mathbf{x})B^i(\mathbf{x})] \quad . \quad (23)$$

This is the required result, as can be seen from evaluating right-hand side of equation below

$$[J^i, P^j] = i\hbar \sum_k \varepsilon^{ijk} P^k = i\hbar \sum_{k,m,n} \varepsilon^{ijk} \varepsilon^{kmn} \varepsilon_0 \int d^3x \, E^m(\mathbf{x}) B^n(\mathbf{x}) \quad (24)$$

with

$$\sum_\kappa \varepsilon^{ijk} \varepsilon^{kmn} = \delta_{im}\delta_{jn} - \delta_{in}\delta_{jm} \quad (25)$$

to give

$$[J^i, P^j] = i\hbar\varepsilon_0 \sum_{m,n} \int d^3x \, E^m(\mathbf{x}) B^n(\mathbf{x}) (\delta_{im}\delta_{jn} - \delta_{in}\delta_{jm}) \quad (26)$$

which leads to (23). Equation (5) is therefore verified.

## 4. Angular momentum commutators

From (3) we have

$$\mathbf{J} = \varepsilon_0 \int d^3x \, \{\mathbf{E}(\mathbf{x})[\mathbf{x}.\mathbf{B}(\mathbf{x})] - [\mathbf{x}.\mathbf{E}(\mathbf{x})]\mathbf{B}(\mathbf{x})\}$$

$$J^i = \varepsilon_0 \sum_m \int d^3x \, x^m [E^i(\mathbf{x})B^m(\mathbf{x}) - E^m(\mathbf{x})B^i(\mathbf{x})] \qquad (27)$$

$$J^j = \varepsilon_0 \sum_n \int d^3y \, y^n [E^j(\mathbf{y})B^n(\mathbf{y}) - E^n(\mathbf{y})B^j(\mathbf{y})]$$

and so

$$[J^i, J^j] = \varepsilon_0^2 \sum_{m,n} \int d^3x \int d^3y \, x^m y^n [\{E^i(\mathbf{x})B^m(\mathbf{x}) - E^m(\mathbf{x})B^i(\mathbf{x})\}, \{E^j(\mathbf{y})B^n(\mathbf{y}) - E^n(\mathbf{y})B^j(\mathbf{y})\}]$$

(28)

The commutators in the curly brackets become

$$\{[E^i(\mathbf{x})B^m(\mathbf{x}), E^j(\mathbf{y})B^n(\mathbf{y})] - [E^i(\mathbf{x})B^m(\mathbf{x}), E^n(\mathbf{y})B^j(\mathbf{y})]$$
$$-[E^m(\mathbf{x})B^i(\mathbf{x}), E^j(\mathbf{y})B^n(\mathbf{y})] + [E^m(\mathbf{x})B^i(\mathbf{x}), E^n(\mathbf{y})B^j(\mathbf{y})]\} \qquad (29)$$

From equation (8) we find the four commutators in the integrand to be

$$[E^i(\mathbf{x})B^m(\mathbf{x}), E^j(\mathbf{y})B^n(\mathbf{y})] \to E^j(\mathbf{y})[E^i(\mathbf{x}), B^n(\mathbf{y})]B^m(\mathbf{x}) - E^i(\mathbf{x})[E^j(\mathbf{y}), B^m(\mathbf{x})]B^n(\mathbf{y})$$
$$-[E^i(\mathbf{x})B^m(\mathbf{x}), E^n(\mathbf{y})B^j(\mathbf{y})] \to -E^n(\mathbf{y})[E^i(\mathbf{x}), B^j(\mathbf{y})]B^m(\mathbf{x}) + E^i(\mathbf{x})[E^n(\mathbf{y}), B^m(\mathbf{x})]B^j(\mathbf{y})$$
$$-[E^m(\mathbf{x})B^i(\mathbf{x}), E^j(\mathbf{y})B^n(\mathbf{y})] \to -E^j(\mathbf{y})[E^m(\mathbf{x}), B^n(\mathbf{y})]B^i(\mathbf{x}) + E^m(\mathbf{x})[E^j(\mathbf{y}), B^i(\mathbf{x})]B^n(\mathbf{y})$$
$$[E^m(\mathbf{x})B^i(\mathbf{x}), E^n(\mathbf{y})B^j(\mathbf{y})] \to E^n(\mathbf{y})[E^m(\mathbf{x}), B^j(\mathbf{y})]B^i(\mathbf{x}) - E^m(\mathbf{x})[E^n(\mathbf{y}), B^i(\mathbf{x})]B^j(\mathbf{y}) \qquad (30)$$

and hence

$$[J^i, J^j] = i\hbar\varepsilon_0 \sum_{m,n,s} \int d^3x \int d^3y \, x^m y^n \frac{\partial}{\partial x^s} \delta(\mathbf{x} - \mathbf{y})$$
$$\{\varepsilon^{ins}[E^m(\mathbf{x})B^j(\mathbf{y}) - E^j(\mathbf{y})B^m(\mathbf{x})] + \varepsilon^{jms}[E^n(\mathbf{y})B^i(\mathbf{x}) - E^i(\mathbf{x})B^n(\mathbf{y})]$$
$$+\varepsilon^{ijs}[E^n(\mathbf{y})B^m(\mathbf{x}) - E^m(\mathbf{x})B^n(\mathbf{y})] + \varepsilon^{mns}[E^j(\mathbf{y})B^i(\mathbf{x}) - E^i(\mathbf{x})B^j(\mathbf{y})]\} \qquad (31)$$

From (15), the eight terms above give, as integrands of $d^3x$ below, with $\mathbf{y}$ set equal to $\mathbf{x}$, the table:

$$+\varepsilon^{ins}\{x^m E^m(\mathbf{x})\}[y^n B^j(\mathbf{y})] \to +\varepsilon^{ins}\{x^m E^m(\mathbf{x})\}[\delta_{sn}+y^n\frac{\partial}{\partial y^s}]B^j(\mathbf{y})$$

$$-\varepsilon^{ins}[y^n E^j(\mathbf{y})]\{x^m B^m(\mathbf{x})\} \to -\varepsilon^{ins}[\delta_{sn}+y^n\frac{\partial}{\partial y^s}]E^j(\mathbf{y})\{x^m B^m(\mathbf{x})\}$$

$$+\varepsilon^{jms}\{y^n E^n(\mathbf{y})\}[x^m B^i(\mathbf{x})] \to -\varepsilon^{jms}\{y^n E^n(\mathbf{y})\}[\delta_{sm}+x^m\frac{\partial}{\partial x^s}]B^i(\mathbf{x})$$

$$-\varepsilon^{jms}[x^m E^i(\mathbf{x})]\{y^n B^n(\mathbf{y})\} \to +\varepsilon^{jms}[\delta_{sm}+x^m\frac{\partial}{\partial x^s}]E^i(\mathbf{x})\{y^n B^n(\mathbf{y})\}$$

$$+\varepsilon^{ijs}[y^n E^n(\mathbf{y})]\{x^m B^m(\mathbf{x})\} \to +\varepsilon^{ijs}[\delta_{sn}+y^n\frac{\partial}{\partial y^s}]E^n(\mathbf{y})\{x^m B^m(\mathbf{x})\}$$

$$-\varepsilon^{ijs}\{x^m E^m(\mathbf{x})\}[y^n B^n(\mathbf{y})] \to -\varepsilon^{ijs}\{x^m E^m(\mathbf{x})\}[\delta_{sn}+y^n\frac{\partial}{\partial y^s}]B^n(\mathbf{y})$$

$$+\varepsilon^{mns}[y^n E^j(\mathbf{y})]\{x^m B^i(\mathbf{x})\} \to +\varepsilon^{mns}[\delta_{sn}+y^n\frac{\partial}{\partial y^s}]E^j(\mathbf{y})\{x^m B^i(\mathbf{x})\}$$

$$-\varepsilon^{mns}\{x^m E^i(\mathbf{x})\}[y^n B^j(\mathbf{y})] \to -\varepsilon^{mns}\{x^m E^i(\mathbf{x})\}[\delta_{sn}+y^n\frac{\partial}{\partial y^s}]B^j(\mathbf{y})$$

. (32)

The first terms in all rows of column 2, except for rows 5 and 6, vanish from the presence of the εδ product. The second terms of column 2 rows 7 and 8 are zero because if *m* and *n* are exchanged a minus results from the ε. It is shown below that second terms of column 2 rows 1 to 6 give zero.

In (32) the second terms of rows 7 and 8 vanish as, if *m* and *n* are interchanged, the ε gives a sign reversal. Finally, the three terms that multiply **x.B**, coming from rows 2, 4 and 5, are

$$[-(\mathbf{x}\times\nabla)E^j |^i +(\mathbf{x}\times\nabla)E^i |^j +\varepsilon^{ijs}\mathbf{x}.\partial\mathbf{E}/\partial x^s]$$ . (33)

Set *i* = *x* and *j* = *y* to get

$$[-(\mathbf{x}\times\nabla)E^y |^x +(\mathbf{x}\times\nabla)E^x |^y +\mathbf{x}.\partial\mathbf{E}/\partial x^z] = z\nabla.\mathbf{E}$$ . (34)

This term is zero because the charge density (= $\varepsilon_0 \mathrm{div}\mathbf{E}$) is zero. The same applies to the other terms by cycling *x, y, z*, and also to the coefficients of **x.E** because $\mathrm{div}\mathbf{B} = 0$. The remaining terms, which are the first terms in rows 5 and 6 of column 2, provide the right hand side of equation (35), which is the result required to satisfy equation (6):

$$[J^i, J^j] = i\hbar\sum_n \varepsilon^{ijn}J^n = i\hbar\varepsilon_0 \sum_{m,n}\varepsilon^{ijn}\int d^3x\, x^m[E^n(\mathbf{x})B^m(\mathbf{x}) - E^m(\mathbf{x})B^n(\mathbf{x})]$$ . (35)

## 5. Conclusion
It has been shown that the classical expressions for the linear and angular momentum of the electromagnetic field equations (2) and (3), satisfy the canonical commutation relations for momentum and angular momentum (4-6) when the gauge-invariant field commutation relations of Bohr and Rosenfeld [6] (1) are used. This confirms their validity as operators that can be used in quantum mechanical calculations of angular momentum [8].

## Appendix A.

We prove here that the order of the of the operators **E** and **B** is not relevant for the quantum operator for the momentum density $\mathbf{p} = \varepsilon_0 \mathbf{E}(\mathbf{x}) \times \mathbf{B}(\mathbf{x})$, expanding on a proof given previously [1]. The second line of (36) below gives the momentum for the standard order of operators **E**×**B**, the third line that with the reversed order - **B**×**E**

$$\mathbf{p}/\varepsilon_0 = \mathbf{E}(\mathbf{x}) \times \mathbf{B}(\mathbf{x}) = \int d^3 y\, \mathbf{E}(\mathbf{x}) \times \mathbf{B}(\mathbf{y}) \delta(\mathbf{x} - \mathbf{y})$$
$$p^i / \varepsilon_0 = \sum_{k,l} \varepsilon^{ikl} \int d^3 y\, E^k(\mathbf{x}) B^l(\mathbf{y}) \delta(\mathbf{x} - \mathbf{y}) \qquad (36)$$
$$p^{i\,\prime}/\varepsilon_0 = \sum_{k,l} \varepsilon^{ikl} \int d^3 y\, B^l(\mathbf{y}) E^k(\mathbf{x}) \delta(\mathbf{x} - \mathbf{y})$$

Note the order of the indices in the Kronecker delta after changing order of the operators. For the order of the operators to be immaterial it is necessary for the difference of the terms below to be zero

$$(p^i - p^{i\,\prime})/\varepsilon_0 = \sum_{k,l} \varepsilon^{ikl} \int d^3 y\, \delta(\mathbf{x} - \mathbf{y}) [E^k(\mathbf{x}), B^l(\mathbf{y})] \qquad (37)$$

If we put in the commutator (1) we get

$$(p^i - p^{i\,\prime})/\varepsilon_0 = -\frac{i\hbar}{\varepsilon_0} \sum_{k,l,s} \varepsilon^{ikl} \varepsilon^{kls} \int d^3 y\, \delta(\mathbf{x} - \mathbf{y}) \frac{\partial}{\partial x^s} \delta(\mathbf{x} - \mathbf{y}) \qquad (38)$$

Using the relation $\varepsilon^{ikl} \varepsilon^{kls} = \varepsilon^{kli} \varepsilon^{kls} = \delta_{ll}\delta_{is} - \delta_{ls}\delta_{il} \to 2\delta_{is}$, when summed over $k$, we have

$$(p^i - p^{i\,\prime})/\varepsilon_0 = -\frac{2i\hbar}{\varepsilon_0} \int d^3 y\, \delta(\mathbf{x} - \mathbf{y}) \frac{\partial}{\partial x^i} \delta(\mathbf{x} - \mathbf{y}) \qquad (39)$$

Taking the three-dimensional delta function to be of the form

$$\delta(\mathbf{r}) = \delta(z)\delta(y)\delta(x) \qquad (40)$$

we get (formally)

$$\int d^3 r\, \delta(\mathbf{r}) \frac{\partial}{\partial x} \delta(\mathbf{r}) = \int_{-\infty}^{\infty} dz\, \delta(z)^2 \int_{-\infty}^{\infty} dy\, \delta(y)^2 \int_{-\infty}^{\infty} dx\, \delta(x) \frac{\partial}{\partial x} \delta(x) \qquad (41)$$

where $x$ (= $x^i$) is one of the coordinates. Of course, the square of a delta function is not defined, so we take the delta functions $\delta(x)$ above to be the limit as a parameter $a$ approaches zero of an even function of $x$, say, a rectangle of width $a$ and area unity. The integral over $x$ is zero by symmetry. The product of the $y$ and $z$ integrals is $1/a^2$. So equation (39) is zero for any non-zero $a$. Provided that $a$ remains finite but small enough so that the variation of the field over a distance *a* is negligible, the integral will still be zero. The proof will therefore not hold for singularities of the field, which give infinitely rapid variation of the fields.

### Appendix B.

In the course of quantizing the free electromagnetic field in the Coulomb gauge with a plane wave basis [3,4] it is necessary to use the geometric relation in {*x, y, z*} space

$$\hat{\varepsilon}_1(\mathbf{k})^i \hat{\varepsilon}_1(\mathbf{k})^j + \hat{\varepsilon}_2(\mathbf{k})^i \hat{\varepsilon}_2(\mathbf{k})^j + \hat{k}^i \hat{k}^j = \delta_{ij} \qquad (42)$$

The $\hat{k}^i$ are the (unit) Cartesian components of the photon wave vector **k**, and $\hat{\boldsymbol{\varepsilon}}_1(\mathbf{k})$ and $\hat{\boldsymbol{\varepsilon}}_2(\mathbf{k})$ are the two photon polarisation unit vectors. These three vectors are orthonormal. Equations (42) are needed to obtain the equal-time commutation relations in the Coulomb gauge for the vector potential operator and its conjugate momentum. These commutation relations have the notable feature of being non-local in space [1,3,4].

The most general explicit form of the vectors $\hat{\boldsymbol{\varepsilon}}_1(\mathbf{k})$ and $\hat{\boldsymbol{\varepsilon}}_2(\mathbf{k})$ does not seem to be available in the quantum electrodynamics literature. The closest to it has been given for the case of rotation angle (see below) $\zeta = 0$ by Loudon [4]. We give the most general form of the vectors here. This general form is given by expressing the three vectors explicitly in Cartesian coordinates {*x, y, z*}:

$$\hat{\mathbf{k}} = \{\sin\theta\cos\phi, \sin\theta\sin\phi, \cos\theta\} \qquad (43)$$

$$\hat{\boldsymbol{\varepsilon}}_1(\mathbf{k}) = \{\cos\theta\cos\phi\sin\zeta + \sin\phi\cos\zeta, -\cos\phi\cos\zeta + \cos\theta\sin\phi\sin\zeta, -\sin\theta\sin\zeta\} \qquad (44)$$

$$\hat{\boldsymbol{\varepsilon}}_2(\mathbf{k}) = \{\cos\theta\cos\phi\cos\zeta - \sin\phi\sin\zeta, \cos\phi\sin\zeta + \cos\theta\sin\phi\cos\zeta, -\sin\theta\cos\zeta\} \qquad (45)$$

The angles $\theta$ and $\phi$ are the spherical angular coordinates of **k**; $\zeta$ is the angle about which the two polarisation vectors are rotated about the **k** axis starting with $\hat{\boldsymbol{\varepsilon}}_1(\mathbf{k})$ in the *x-y* plane and $\hat{\boldsymbol{\varepsilon}}_2(\mathbf{k})$ in a plane containing the *z* axis; this occurs at $\zeta = 0$. Equations (43-45) describe the most general configuration of the three vectors. It is straightforward to verify that they satisfy equations (42) and the orthonormality conditions for all *i, j, θ, φ* and *ζ*.

We comment on the derivation of the new expressions (44-45). These vectors have been given by Loudon [4] for the special case of $\zeta = 0$. By using the Rodrigues rotation formula [9]

$$\mathbf{x}' = \mathbf{x}\cos\zeta + (\hat{\mathbf{k}} \times \mathbf{x})\sin\zeta + \hat{\mathbf{k}}(\hat{\mathbf{k}}.\mathbf{x})(1-\cos\zeta) \qquad (46)$$

which transforms the vector **x** to a vector **x'** by rotation around an axis **k** by the angle $\zeta$, equations (44-45) are obtained from the $\zeta = 0$ result.

**Acknowledgement.**

**References.**